\documentclass[aps,prl, superscriptaddress, twocolumn]{revtex4}
\usepackage{bbm}
\usepackage{amssymb}
\usepackage{amsmath}
\usepackage{verbatim}
\usepackage[pdftex]{graphicx}
\usepackage{dsfont}
\usepackage{color}

\begin{document}

\title{Braiding of Atomic Majorana Fermions in Wire Networks and Implementation of the Deutsch-Josza Algorithm}
\author{Christina V. Kraus}
\affiliation{Institute for Quantum Optics and Quantum Information of the Austrian Academy of Sciences, A-6020 Innsbruck, Austria }
\affiliation{Institute for Theoretical Physics, Innsbruck University, A-6020 Innsbruck, Austria}
\author{P. Zoller}
\affiliation{Institute for Quantum Optics and Quantum Information of the Austrian Academy of Sciences, A-6020 Innsbruck, Austria }
\affiliation{Institute for Theoretical Physics, Innsbruck University, A-6020 Innsbruck, Austria}
\author{Mikhail A. Baranov}
\affiliation{Institute for Quantum Optics and Quantum Information of the Austrian Academy of Sciences, A-6020 Innsbruck, Austria }
\affiliation{Institute for Theoretical Physics, Innsbruck University, A-6020 Innsbruck, Austria}
\affiliation{RRC "Kurchatov Institute", Kurchatov Square 1, 123182, Moscow, Russia}

\begin{abstract}
We propose an efficient protocol for braiding atomic Majorana fermions in wire networks with AMO techniques
and demonstrate its robustness against experimentally relevant errors. Based on this protocol we provide a topologically protected implementation 
of the Deutsch-Josza algorithm.
\end{abstract}

\maketitle

\address{Institute for Theoretical Physics, University of Innsbruck, A-6020
Innsbruck, Austria}

The prediction of particles with anyonic statistics in topological phases of matter has resulted in the proposal of decoherence-free Topological Quantum Computation (TQC) \cite{TQC_Kitaev,NayakRMP, Pachos}. TQC requires the creation of anyonic particles as well as their controlled interchange, known as \emph{braiding}, which is the fundamental building block of topological quantum gates \cite{DasSarma_TQC1, DasSarma_TQC2}. While the implementation of these tasks in real physical systems is an outstanding challenge, the reported observation of anyonic Majorana fermions (MFs) in hybrid superconductor-semiconductor nanowire devices \cite{MajoranaExperiment1, MajoranaExperiment2, MajoranaExperiment3} and the proposals for the manipulation \cite{Alicea_TQC, OregHalperin, Akhmerov} of anyonic Majorana fermions (MFs) in solid state systems are promising first steps in this direction \cite{Beenakker_review, Hassler, Akhmerov, Flensberg1, Flensberg2, Demler}.  A complementary and promising approach towards realizing and coherently control MFs are ultracold atoms confined to 1D optical lattices coupled to BCS or molecular atomic reservoirs. The recent realization of a quantum gas microscope \cite{Greiner_singlesite, SinglesiteMicroscope} for optical lattices adds single site addressing and measurement to the toolbox of possible atomic operations to create and detect MFs \cite{Probing_Liang, MajoranaDetection, Nascimbene}.

Building on these experimental advances, we describe in this Letter an efficient braiding protocol for atomic MFs, based on performing simple lattice operations on a few sites in an array of 1D wires, and we provide a careful study of the full braiding dynamics including imperfections. In addition, we will show that these elementary braiding operations, although they do not represent the complete set of quantum gates \cite{NayakBasis, Werner}, can be combined to realize a Deutsch-Jozsa algorithm \cite{DeutschJosza}, demonstrating that the implementation of simple quantum algorithms in atomic topological setups is within experimental reach.

\begin{figure}
\begin{center}
\includegraphics[width = 0.7 \columnwidth]{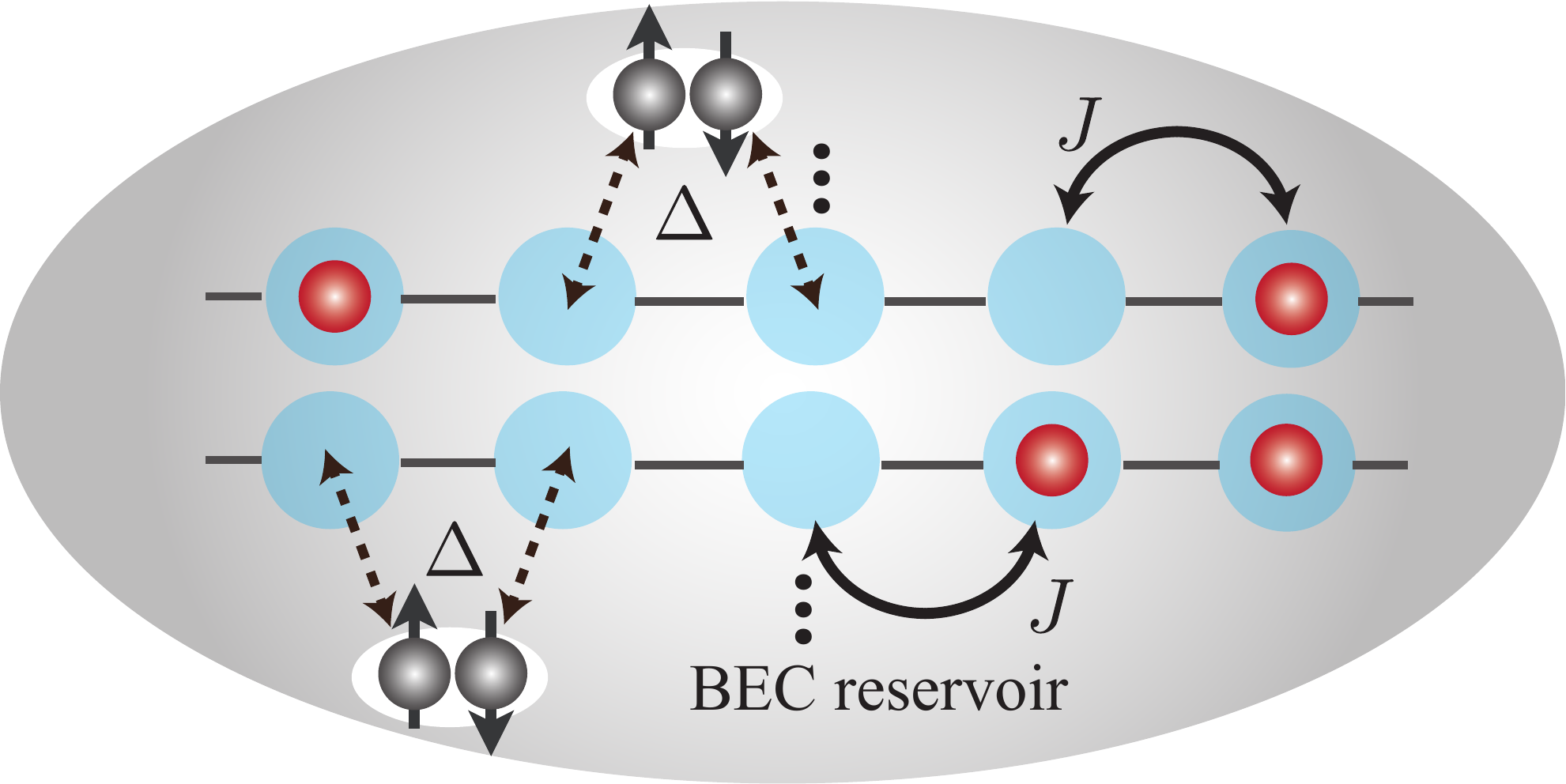}
\caption{Realization of an array of one-dimensional Kitaev wires in an optical lattice setup: Atoms (red circles) can hop between neighboring sites (blue circles) with strength $J$ along the individual wires.  The pairing term of strength $\Delta$ can be realized by a Raman
induced dissociation of Cooper pairs (or Feshbach molecules) forming an
atomic BCS reservoir.\label{fig:wiresetup}}
\end{center}
\end{figure}
\emph{Braiding of atomic Majorana fermions.} We consider a system of single
component fermions that are confined to an array of one-dimensional ($1D$)
wires of $L$ sites (see Fig.~\ref{fig:wiresetup}) and that are governed by a Hamiltonian $H =
\sum_{n} H^{(n)}$. The Hamiltonian $H^{(n)} =
\sum_{j=1}^{L-1}-Ja^{\dagger}_{n,j}a_{n,j+1} + \Delta a_{n,j}a_{n,j+1} +
h.c. - \mu \sum_n a^{\dagger}_n a_n$ realizes a Kitaev chain \cite{Kitaevchain} in the $n$-th
wire. The operators $a^{\dagger}_{n,j}$ and $a_{n,j}$ are fermionic creation
and annihilation operators, $J>0$ and $\Delta \in \mathds{R}$ are
nearest-neighbor hopping and pairing terms, and $\mu$ is a chemical
potential. As demonstrated in \cite{Probing_Liang}, a Hamiltonian of the form $H^{(n)}$ allows
for a cold atom implementation: While the hopping term arises naturally in
an optical lattice setup, the pairing term can be realized by a Raman
induced dissociation of Cooper pairs (or Feshbach molecules) forming an
atomic BCS reservoir.

It has been shown in \cite{Kitaevchain} that the Hamiltonian $H^{(n)}$ supports zero energy
Majorana fermions of the form $\gamma
_{L/R}^{(n)}=\sum_{j}v_{n,j}^{L/R}c_{n,j}$ with (real) coefficients $%
v^{L/R}_{n,j}$ which are localized at the left/right end of the $n$-th
wire. Here, $c_{n,2j-1}=a_{n,j}^{\dagger }+a_{n,j}$ and $%
c_{n,2j}=(-i)(a_{n,j}^{\dagger }-a_{n,j})$ are Majorana operators fulfilling 
$\{c_{n,k},c_{m,l}\}=2\delta _{kl}\delta _{mn}$. For the "ideal" quantum
wire $(J=|\Delta |,\mu =0)$, one has $v_{n,1}^{L}=1,v_{n,2L}^{R}=1$ and else 
$v_{n,j}^{L/R}=0$. Otherwise, the modes $\gamma _{L/R}^{(n)}$ decay
exponentially inside the bulk. Each wire has two degenerate ground states $%
|0_{n}\rangle $ and $|1_{n}\rangle $ with even and odd parity, respectively,
corresponding to the presence or absence of the Majorana fermion $%
f_{n}=\gamma _{L}^{(n)}-i\gamma _{R}^{(n)}$, i.e. $f_{n}|0_{n}\rangle =0$, $%
f_{n}^{\dagger }|0_{n}\rangle =|1_{n}\rangle $. For a proposal how to
prepare MFs in the desired parity subspace see \cite{MajoranaDetection}.

\begin{figure}[tbp]
\begin{center}
\includegraphics[width=0.95\columnwidth]{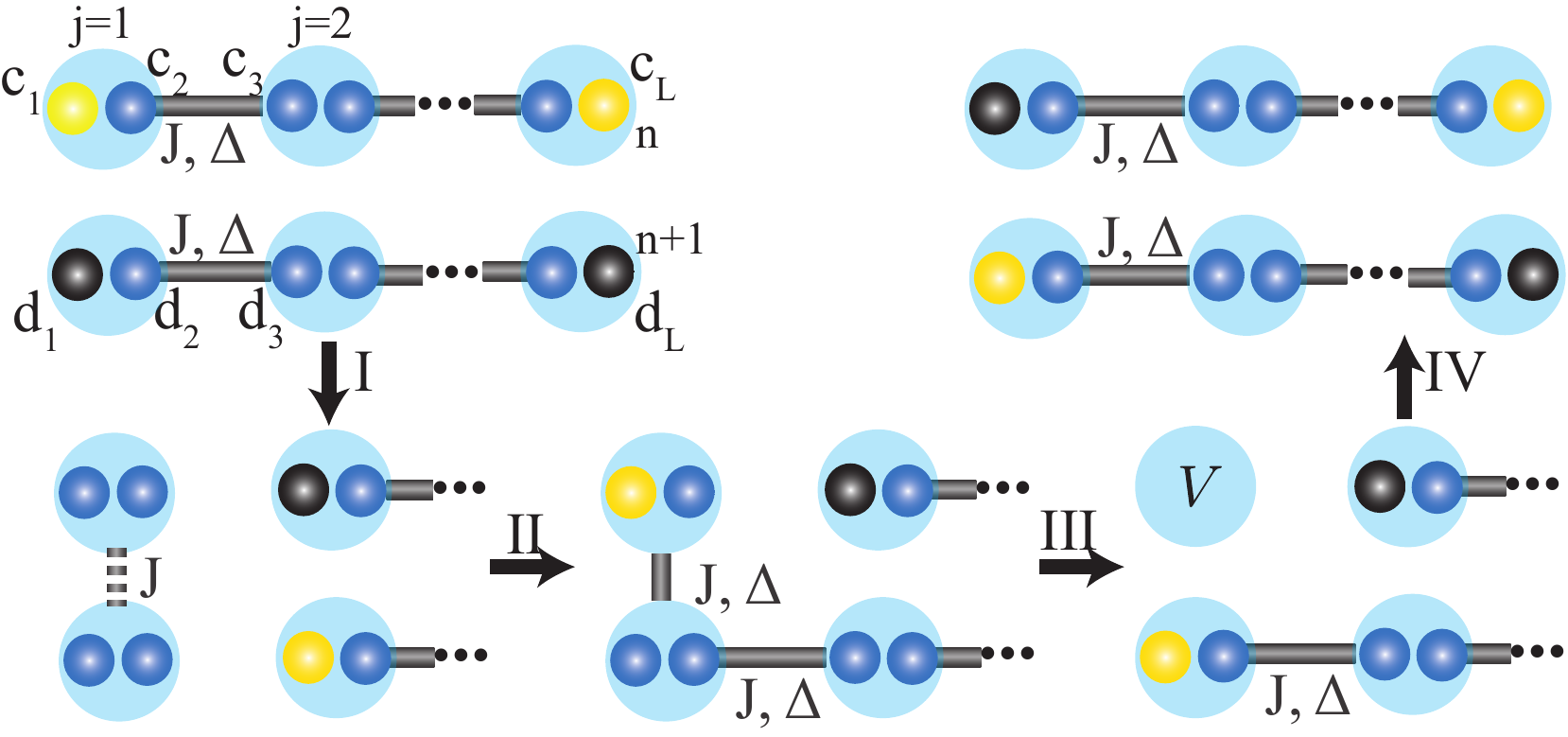}
\end{center}
\caption{Braiding protocol for two perfect quantum wires. The zero-energy
Majorana modes that are initially on the upper (lower) wire are shown as
yellow (black) spheres, while the blue ones corresponds to the Majorana
operators which are coupled into finite-energy fermionic modes. Coupling of
Majorana operators via hopping and pairing (Kitaev coupling) is indicated by
grey solid links, while the coupling via hopping only is shown as a
dashed link.}
\label{fig:Braiding}
\end{figure}

Since MFs exhibit anyonic statistics, an appropriate interchange of two
Majorana modes, $\gamma _{1,2}$, allows to realize the braiding unitary $%
U_{b}=e^{\pi \gamma _{1}\gamma _{2}/4}$. This interchange which leads to the
transformation $\gamma _{1}\mapsto -\gamma _{2}$, $\gamma _{2}\mapsto \gamma
_{1}$ resulting in a non-trivial phase factor for the wave function is the key
step for realizing a TQC. In the following we present a protocol for a cold
atom implementation that allows to realize braiding. To this end, we
consider two neighboring wires $n$ and $n+1$ governed by two ideal Kitaev
Hamiltonians $H^{(n)}$ and $H^{(n+1)}$. The use of ideal wires allows for a
simple analytic treatment because only six Majorana operators are involved
in the protocol. It is convenient to label the sites by $(w,j)$, where $%
w=u,l $ denotes the upper $(n)$ resp. lower $(n+1)$ wire and $j=1,\ldots ,L$%
. We label the sites that are involved in our protocol as $\vec{s}_{1}=(u,1)$%
, $\vec{s}_{2}=(u,2)$,$\vec{s}_{3}=(l,1)$ and $\vec{s}_{4}=(l,2)$ (see Fig.~%
\ref{fig:Braiding}). To simplify notation, we write $c_{u,j}\equiv c_{j}$,  and $c_{l,j}\equiv d_{j}$ with
Majorana modes $\gamma _{L}^{(u)}=c_{1}$, $\gamma _{R}^{(u)}=c_{L}$, $\gamma
_{L}^{(l)}=d_{1}$, $\gamma _{R}^{(l)}=d_{L}$.

Let us now show how to braid the left Majorana modes $\gamma _{L}^{(u)}$ and 
$\gamma _{L}^{(l)}$ around each other with only local (adiabatic) changes in
the Hamiltonian on the left edge of the system. These changes include
switching on/off (i) the hopping $H_{\vec{s}_{i},\vec{s}_{j}}^{(h)}=-Ja_{%
\vec{s}_{i}}^{\dagger }a_{\vec{s}_{j}}+h.c.$ and (ii) the pairing $H_{\vec{s}%
_{i},\vec{s}_{j}}^{(p)}=Ja_{\vec{s}_{i}}a_{\vec{s}_{j}}+h.c.$ between the
neighboring sites $\vec{s}_{i}$ and $\vec{s}_{j}$, and (iii) the local
potential $H_{\vec{s}_{i}}^{({lp)}}=2Va_{\vec{s}_{i}}^{\dagger }a_{\vec{s}%
_{i}}$ on site $\vec{s}_{i}$. Note that a combination of (i) and (ii) allows
to switch on/off the Kitaev coupling $H_{\vec{s}_{i},\vec{s}_{j}}^{(K)}=H_{%
\vec{s}_{i},\vec{s}_{j}}^{(h)}+H_{\vec{s}_{i},\vec{s}_{j}}^{(p)}$. These
operations are based on the single site/link addressing available in cold
atom experiments \cite{Bloch_singlespin,Greiner_singlesite}.

Let us now describe the braiding protocol in detail. The physical process
behind is the transfer of one fermion from the system (i. e. either from the
upper or from the lower wire) into the lower wire. We characterize the
required adiabatic changes via a time-dependent parameter $\phi _{t}$ that
varies from $0$ to $\pi /2$, and perform them in four steps. In describing
these steps, we will only write down the Hamiltonian for the four involved
sites and follow the evolution of the zero modes which are always separated
by a finite gap from the rest of the spectrum. 

\emph{Step I:} We decouple the two very left sites, $\vec{s}_{1}$ and $\vec{s%
}_{3}$ from the system by switching off the couplings $H_{\vec{s}_{i},\vec{s}%
_{j}}^{(K)}$ between sites $\vec{s}_{1}-\vec{s}_{2}$ and $\vec{s}_{3}-\vec{s}%
_{4}$, and, at the same time, switch on the hopping between sites $\vec{s}%
_{1}-\vec{s}_{3}$: 
\begin{align*}
& H_{I}(t)=\cos \phi _{t}(H_{\vec{s}_{1}\vec{s}_{2}}^{(K)}+H_{\vec{s}_{3}%
\vec{s}_{4}}^{(K)})+\sin \phi _{t}H_{\vec{s}_{1}\vec{s}_{3}}^{(h)} \\
& =-iJ\left[ \cos \phi _{t}(c_{2}c_{3}+d_{2}d_{3})+\sin \phi
_{t}(c_{2}d_{1}-c_{1}d_{2})/2\right] .
\end{align*}%
During this process the zero modes evolve according to $\gamma
_{L}^{(u)}(\phi _{t})=(2\cos \phi _{t}c_{1}-\sin \phi _{t}d_{3})/\sqrt{1+3\cos^2 \phi_t}$, $\gamma
_{L}^{(l)}(\phi _{t})=(2\cos \phi _{t}d_{1}-\sin \phi _{t}c_{3})/\sqrt{1+3\cos ^2 \phi_t}$, such that at
the end $\gamma _{L}^{(u)}=-d_{3}$ and $\gamma _{L}^{(d)}=-c_{3}$. Note that
the two decoupled sites $\vec{s}_{1}$ and $\vec{s}_{3}$ carry exactly one
fermion which has been taken out of the system.\newline
\emph{Step II:} We put now this fermion in the lower wire by switching on $%
H_{\vec{s}_{i},\vec{s}_{j}}^{(K)}$ between sites $\vec{s}_{3}-\vec{s}_{4}$,
and $H_{\vec{s}_{i},\vec{s}_{j}}^{(p)}$ between the sites $\vec{s}_{1}-\vec{s%
}_{3}$: 
\begin{align*}
& H_{II}(t)=H_{\vec{s}_{1}\vec{s}_{3}}^{(h)}+\sin \phi _{t}\left( H_{\vec{s}%
_{1}\vec{s}_{3}}^{(p)}+H_{\vec{s}_{3}\vec{s}_{4}}^{(K)}\right) = \\
& -i\tfrac{J}{2}\left[ (c_{2}d_{1}-c_{1}d_{2})+\sin \phi
_{t}(c_{2}d_{1}+c_{1}d_{2}+2d_{2}d_{3})\right] .
\end{align*}%
The zero modes evolve as $\gamma _{L}^{(u)}(\phi _{t})=(2\sin \phi
_{t}c_{1}-(1-\sin \phi _{t})d_{3})/\sqrt{4 \sin^2 \phi_t + (1-\sin\phi_t)^2}$, $\gamma _{L}^{(l)}=-c_{3}$, such that at
the end $\gamma _{L}^{(u)}=c_{1}$ and $\gamma _{L}^{(d)}=-c_{3}$. Note, that
at this stage the Majorana mode $\gamma _{L}^{(u)}$ ($\gamma _{L}^{(l)}$)
has already been moved from the upper (lower) to the lower (upper) wire.
However, two additional steps are needed to recover the original
configuration of the wires. \newline
\emph{Step III:} We move the Majorana mode from the site $\vec{s}_{1}$ to
the site $\vec{s}_{3}$ by switching on $H_{\vec{s}_{1}}^{(lp)}$ and
simultaneously switching off $H_{\vec{s}_{i}\vec{s}_{j}}^{(K)}$ between the
sites $\vec{s}_{1}-\vec{s}_{3}$: 
\begin{align*}
H_{III}(t)& =\sin \phi _{t}H_{\vec{s}_{1}}^{(lp)}+\cos \phi _{t}H_{\vec{s}%
_{1}\vec{s}_{3}}^{(K)}+H_{\vec{s}_{3}\vec{s}_{4}}^{(K)} \\
& =-iJ(c_{2}d_{1}\cos \phi _{t}+d_{2}d_{3})-iV\sin \phi _{t}c_{1}c_{2}.
\end{align*}%
The evolution of the zero mode $\gamma _{L}^{(u)}=(J\cos \phi
_{t}c_{1}+V\sin \phi _{t}d_{1})/\sqrt{(J\cos \phi _{t})^{2}+(V\sin \phi
_{t})^{2}}$ results in $\gamma _{L}^{(u)}=d_{1}$, while $\gamma
_{L}^{(l)}=-c_{3}$ remains fixed.\newline
\emph{Step IV:} Finally, we switch off $H_{\vec{s}_{1}}^{(lp)}$ and switch
on $H_{\vec{s}_{1}\vec{s}_{2}}^{(K)}$: 
\begin{align*}
H_{IV}(t)& =\sin \phi _{t}H_{\vec{s}_{1}\vec{s}_{2}}^{(K)}+H_{\vec{s}_{3}%
\vec{s}_{4}}^{(K)}+\cos \phi _{t}H_{\vec{s}_{1}}^{(lp)} \\
& =-iJ\left[ \sin \phi _{t}c_{2}c_{3}+d_{2}d_{3}\right] -iV\cos \phi
_{t}c_{1}c_{2}
\end{align*}%
The zero modes are given by $\gamma _{L}^{(u)}=d_{1}$, $\gamma
_{L}^{(l)}=-(J\sin \phi _{t}c_{1}+V\cos \phi _{t}c_{3})/\sqrt{(J\sin \phi
_{t})^{2}+(V\cos \phi _{t})^{2}}$, so that finally we get the desired
braiding $\gamma _{L}^{(u)}\mapsto \gamma _{L}^{(l)}$ and $\gamma
_{L}^{(l)}\mapsto -\gamma _{L}^{(u)}$ of left the Majorana modes on the wires $n$
and $n+1$, which corresponds (up to unimportant phase factor) to the unitary 
$U_{n}=e^{\pi \gamma _{L}^{(u)}\gamma _{L}^{(l)}/4}$.

Note that the braiding in the other direction, $U_{n}^{\dagger }$ and $%
\gamma _{L}^{(u)}\mapsto -\gamma _{L}^{(l)}$, $\gamma _{L}^{(l)}\mapsto
\gamma _{L}^{(u)}$, can be achieved by putting the uncoupled fermion in the
upper (instead of the lower) wire with a simple modification of Steps II-IV.

The braiding results in the change of the correlation functions of the
Majorana operators (see Fig. 2) and thus changes also the long-range
fermionic correlations. This can also be translated into the change of the
fermionic parities of the wires: If $|0_{n}\rangle $ ($|1_{n}\rangle $)
denotes the state of the $n$-th wire with even (odd) parity and, for
example, we start from the state $|0_{n}0_{n+1}\rangle $ with both wires
with even parity, then the braiding $U_{n}$ results in $U_{n}|0_{n}0_{n+1}%
\rangle =(|0_{n}0_{n+1}\rangle +|1_{n}1_{n+1}\rangle )/\sqrt{2}$, and $%
U_{n}^{2}|0_{n}0_{n+1}\rangle =|1_{n}1_{n+1}\rangle $. The result of the
braiding, therefore, can be checked by measuring the change of the Majorana
correlation functions in Time-of-Flight or spectroscopic experiments \cite{MajoranaDetection}, or
by measuring the parity of the wires by counting the number of fermions
modulo two \cite{SinglesiteMicroscope}.
\begin{figure}[tbp]
\includegraphics[width=0.95\columnwidth]{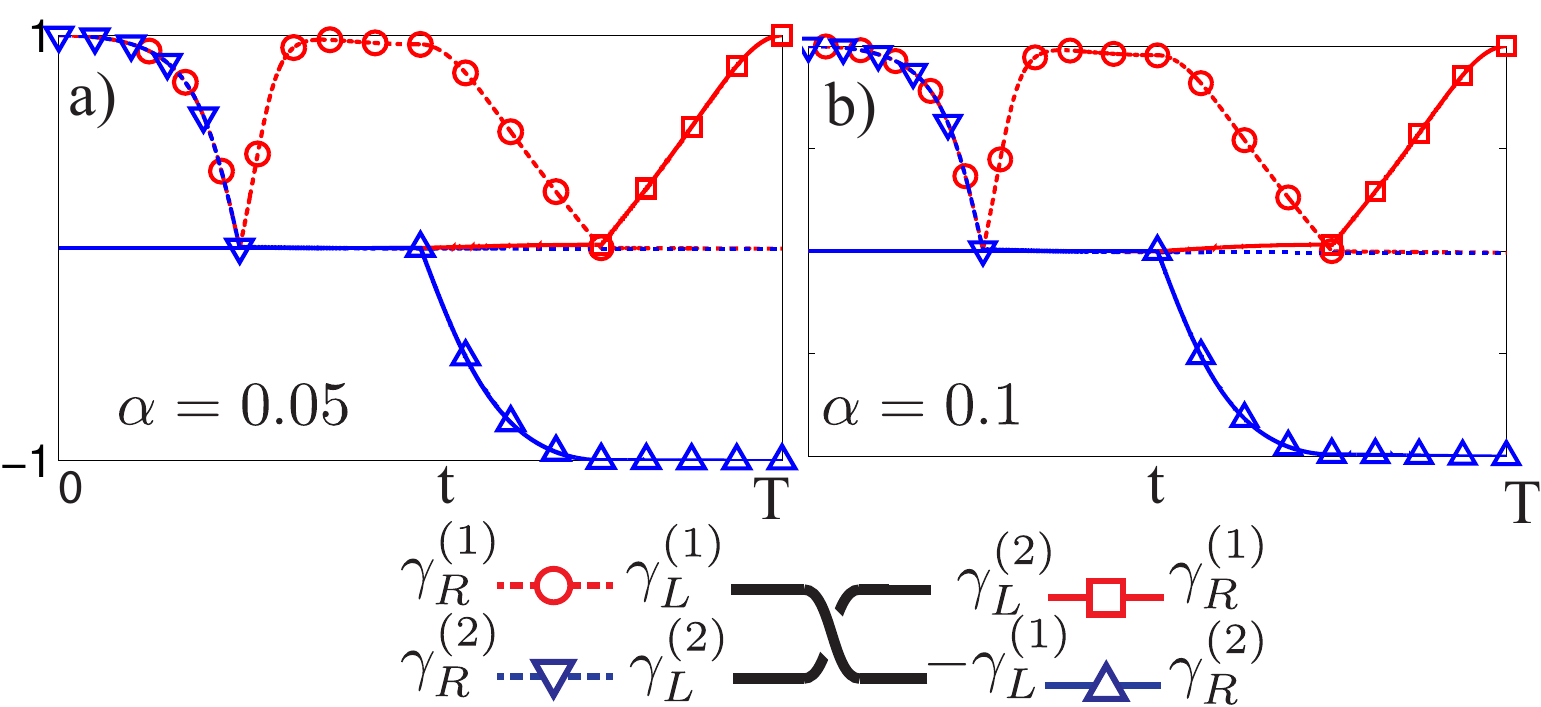}
\caption{Evolution of the Majorana correlation functions $\langle i\protect%
\gamma _{L}^{(1)}\protect\gamma _{R}^{(1)}\rangle $ (red, $\circ$), $\langle i%
\protect\gamma _{L}^{(2)}\protect\gamma _{R}^{(2)}\rangle $ (blue, $\bigtriangledown$), $%
\langle i\protect\gamma _{L}^{(2)}\protect\gamma _{R}^{(1)}\rangle $ (red,
$\square$), and $\langle i\protect\gamma _{L}^{(1)}\protect\gamma %
_{R}^{(2)}\rangle $ (blue,  $\bigtriangleup$) during the braiding protocol with errors $%
\protect\alpha $ in the local operations for two non-ideal quantum wires
with $|\Delta |=1.5J$ and $\protect\mu =0$. Markers are only drawn in regions where the correlation functions are non-zero.}
\label{fig:nonperfect}
\end{figure}

\emph{Non-ideal wires and non-perfect operations.} We have just demonstrated
the braiding for the case of ideal Kitaev wires and perfect local operations
(single site/link addressing). Remarkably, the topological origin of the
Majorana modes ensures the robustness of the results of the braiding
protocol based on Steps I-IV also in the realistic case of non-ideal wires
and local operations provided the Majorana modes are spatially
well-separated. We have checked this numerically by considering two
non-ideal wires with $J\neq |\Delta |$, $\mu \neq 0$ and assuming that the
local operations have an error $\alpha $ in the following sense: (i)
Switching on the hopping $J$ and/or the pairing $\Delta $ between the sites $%
(u,1)-(d,1)$, also introduces the hopping $J\alpha $ and/or the pairing $%
\alpha \Delta $ between the adjacent sites $(u,2)-(d,2)$. (ii) Switching off
the couplings between the sites $(w,1)-(w,2)$ also reduces the couplings
between the sites $(w,2)-(w,3)$ by a factor $(1-\alpha )$. (iii) Raising the
local potential $V$ on the site $(u,1)$ results in a local potential $\alpha
V$ on the neighboring sites $(u,2)$ and $(l,1)$. As an example, we present in Fig.~\ref%
{fig:nonperfect} numerical results of the braiding protocol with
errors $\alpha =0.05$ and $\alpha =0.1$ in the local operations for two
quantum wires of the length $L=40$ with $|\Delta |=1.5J$ and $\mu =0$. One
can clearly see the robustness of the final results of the braiding.

\begin{figure}[t]
\includegraphics[width=0.99\columnwidth]{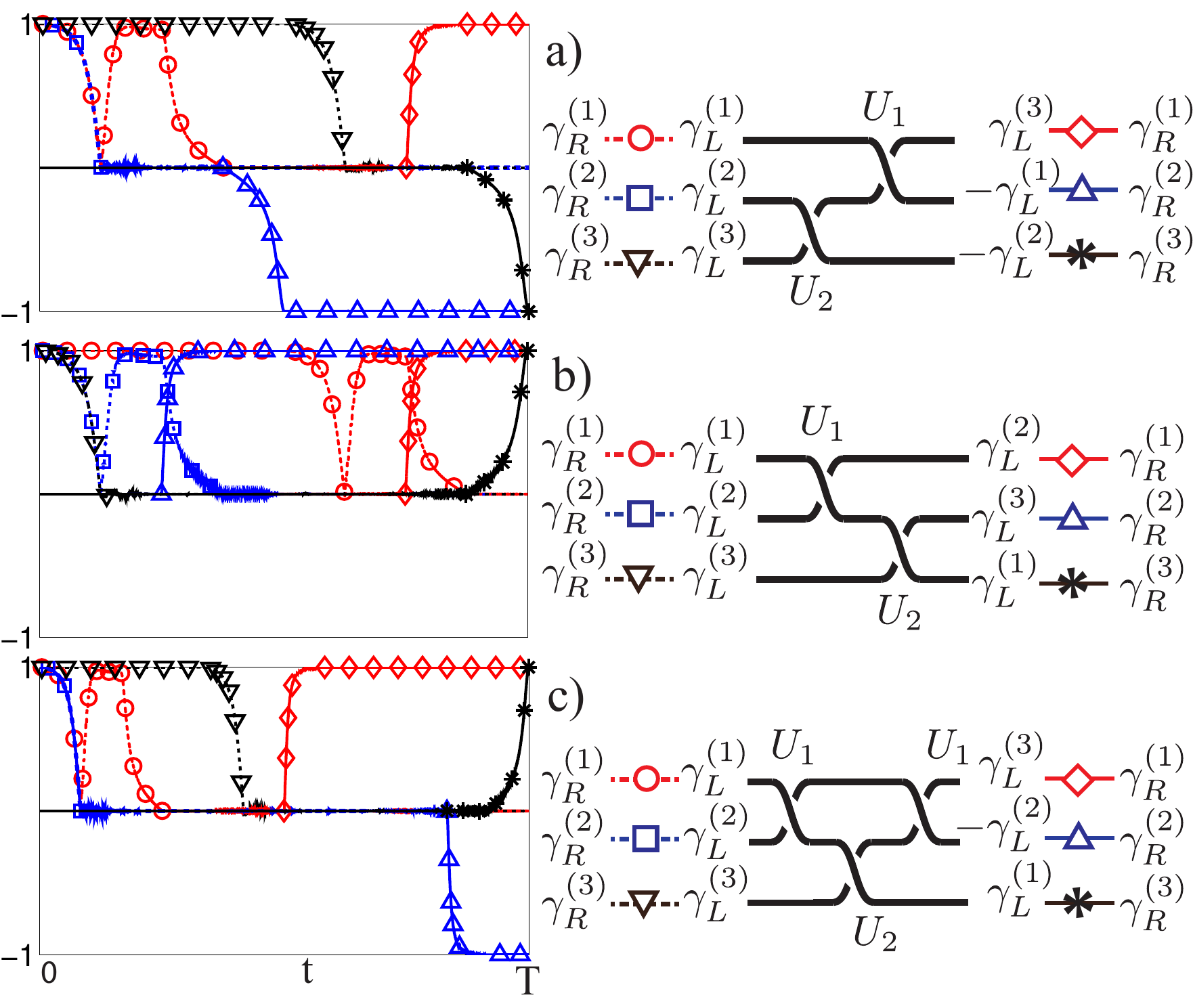}
\caption{Braid group in a setup of three wires. We present the real-time
evolution of the correlations functions $i\langle \protect\gamma _{L}^{(n)}%
\protect\gamma _{R}^{(m)}\rangle $ under the action of (a)$U_{2}U_{1}$, (b) $%
U_{1}U_{2}$ and (c) $U_{1}U_{2}U_{1}$ for a chain of the length $L=40$ with  
$|\Delta |=1.5J$ and $\protect\mu =0$. Markers are only drawn in regions where the correlation functions are non-zero.}
\label{fig:braidgroup_numerics}
\end{figure}
\emph{Braid group.} It is also easy to check that the unitary transformations 
$U_{n}$ of the Majorana operators corresponding to the braiding protocol
fullfill all necessary conditions of the braid group~\cite{NayakRMP}: For any two braiding
unitaries $U_{n}$ and $U_{n+1}$, one has $U_{n}U_{n+1}\neq U_{n+1}U_{n}$ and 
$U_{n-1}U_{n}U_{n-1}=U_{n}U_{n-1}U_{n}$.

To demonstrate this, consider three wires with left Majorana modes $\gamma
_{L}^{(1)}$, $\gamma _{L}^{(2)}$ and $\gamma _{L}^{(3)}$, and braiding
unitaries $U_{1}=e^{\pi \gamma _{L}^{(1)}\gamma _{L}^{(2)}/4}$ and $%
U_{2}=e^{\pi \gamma _{L}^{(2)}\gamma _{L}^{(3)}/4}$ that braid the modes $%
\gamma _{L}^{(1)}$, $\gamma _{L}^{(2)}$ and $\gamma _{L}^{(2)}$, $\gamma
_{L}^{(3)}$, respectively. The braid group conditions for $U_{1}$ and $U_{2}$
then immediately follow from the following formulae 
\begin{align}
& U_{1}U_{2}(\gamma _{L}^{(1)},\gamma _{L}^{(2)},\gamma _{L}^{(3)})=(\gamma
_{L}^{(3)},-\gamma _{L}^{(1)},-\gamma _{L}^{(2)})  \notag \\
& U_{2}U_{1}(\gamma _{L}^{(1)},\gamma _{L}^{(2)},\gamma _{L}^{(3)})=(\gamma
_{L}^{(2)},\gamma _{L}^{(3)},\gamma _{L}^{(1)}) \\
& U_{1}U_{2}U_{1}(\gamma _{L}^{(1)},\gamma _{L}^{(2)},\gamma
_{L}^{(3)})=(\gamma _{L}^{(3)},-\gamma _{L}^{(2)},\gamma _{L}^{(1)})=  \notag
\\
& U_{2}U_{1}U_{2}(\gamma _{L}^{(1)},\gamma _{L}^{(2)},\gamma _{L}^{(3)}).
\end{align}

These properties can be tested experimentally by measuring the corresponding
changes of the fermionic correlation functions. For example, the action of $%
U_{1}U_{2}$ results in $i\langle \gamma _{L}^{(1)}\gamma _{R}^{(1)}\rangle
\mapsto i\langle \gamma _{L}^{(3)}\gamma _{R}^{(1)}\rangle $, $i\langle
\gamma _{L}^{(2)}\gamma _{R}^{(2)}\rangle \mapsto -i\langle \gamma
_{L}^{(1)}\gamma _{R}^{(2)}\rangle $ and $i\langle \gamma _{L}^{(3)}\gamma
_{R}^{(3)}\rangle \mapsto -i\langle \gamma _{L}^{(2)}\gamma
_{R}^{(3)}\rangle $, while $U_{1}U_{2}U_{1}$ produces the following changes: 
$i\langle \gamma _{L}^{(1)}\gamma _{R}^{(1)}\rangle \mapsto i\langle \gamma
_{L}^{(3)}\gamma _{R}^{(1)}\rangle $, $i\langle \gamma _{L}^{(2)}\gamma
_{R}^{(2)}\rangle \mapsto -i\langle \gamma _{L}^{(2)}\gamma
_{R}^{(1)}\rangle $, and $i\langle \gamma _{L}^{(3)}\gamma _{R}^{(3)}\rangle
\mapsto i\langle \gamma _{L}^{(1)}\gamma _{R}^{(3)}\rangle $ (see Fig.~\ref%
{fig:braidgroup_numerics}). This change in the correlation functions can be
measured, for example, in TOF or spectroscopic experiments as proposed in \
Ref.~\cite{MajoranaDetection}.

\emph{Deutsch-Josza algorithm.} Although the braiding of MFs is robust, it
does not provide a tool to construct a universal set of gates needed for
TQC: As it has been shown in Ref.~\cite{Werner}, only a subgroup of the Clifford group
can be realized via braiding. Fortunately, not all QC algorithms require a
universal set of gates. One example is the Deutsch-Josza algorithm \cite{DeutschJosza} which, as
we will show below, can be implemented for two qubits in a remarkably
efficient way via braiding of MFs.

The Deutsch-Josza algorithm allows to determine whether the function
("oracle") $g(x)$ which is defined on the space of states of $n$ qubits and
takes the values $0$ or $1$, $g:\{\left\vert 0\right\rangle ,\left\vert
1\right\rangle \}^{\otimes n}\mapsto \{0,1\}$, is constant (has the same
value, say, $0$, for all inputs) or balanced (takes value $0$ for half of
the inputs, and $1$ for the other half). For the algorithm to work, the
function $g$ has to be implemented as the unitary $U_{g}:|x\rangle \mapsto
(-1)^{g(x)}|x\rangle $, where $|x\rangle \in \{\left\vert 0\right\rangle
,\left\vert 1\right\rangle \}^{\otimes n}$, which is actually a major
problem for experimental realizations: A faulty oracle spoils the quantum speedup \cite{Regev}.

For two qubits with the computational basis $\{|00\rangle ,|01\rangle
,|10\rangle ,|11\rangle \}$, a possible choice for $U_{g}$ is

\begin{align*}
U_{g_{0}}& =\mathrm{diag}(1,1,1,1),\,\,U_{g_{1}}=\mathrm{diag}%
(1,1,-1,-1),\,\, \\
U_{g_{2}}& =\mathrm{diag}(1,-1,-1,1),\,\,U_{g_{3}}=\mathrm{diag}(1,-1,1,-1),
\end{align*}
for the constant $g_{0}$ and the balanced $g_{1}$, $g_{2}$, and $g_{3}$
oracle functions, respectively. (Note that an equivalent set of oracles can
be obtained by multiplying the above unitaries with $-1$.) The algorithm
works then in the following way: After preparing the system in the state $%
|00\rangle $, we apply the Hadamard gate $H$ to each qubit, $H\left\vert
0\right\rangle =(\left\vert 0\right\rangle +\left\vert 1\right\rangle )/%
\sqrt{2}$, $H\left\vert 1\right\rangle =(\left\vert 0\right\rangle
-\left\vert 1\right\rangle )/\sqrt{2}$, then we apply the unitary $U_{g}$
corresponding to the oracle under test, then again the Hadamard gate to each
qubit, and, finally, we measure the probability to find the system in the
state $|00\rangle $. This probability is $1$ if $g(x)$ is constant, and $0$
if $g(x)$ is balanced, as can be seen from the following calculations

\begin{align}
& |00\rangle \overset{H\otimes H}{\mapsto }\frac{1}{2}\sum_{\mathbf{x}}|%
\mathbf{x}\rangle \overset{U_{g}}{\mapsto }\frac{1}{2}\sum_{\mathbf{x}%
}(-1)^{g(\mathbf{x})}|\mathbf{x}\rangle   \notag \\
& \overset{H\otimes H}{\mapsto }\frac{1}{4}\sum_{\mathbf{x}}(-1)^{g(\mathbf{x%
})}\sum_{\mathbf{y}}(-1)^{\mathbf{x}\cdot \mathbf{y}}|\mathbf{y}\rangle ,
\end{align}%
where we define $\mathbf{x}=(x_{1},x_{2})$ and $\mathbf{y}=(y_{1},y_{2})$
with $x_{i}$, $y_{i}\in \{0,1\}$, and $\mathbf{x}\cdot \mathbf{y}%
=x_{1}y_{1}+x_{2}y_{2}$. 
\begin{figure}[tbp]
\includegraphics[width=0.95\columnwidth]{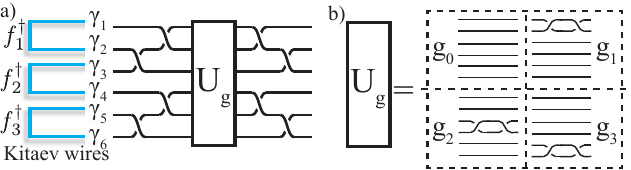}
\caption{a) Setup and implementation of the "oracle"  Deutsch-Josza
algorithm for two qubits via braiding. b) Implementation of the "oracle"
unitary $U_{g}$ via braiding (see main text).\label{fig:Deutsch_Josza} }
\end{figure}

To implement the above algorithm, we use a setup of three quantum wires in
the geometry shown in Fig.~\ref{fig:Deutsch_Josza}  and define a computational basis for two
qubits as $|00\rangle =f_{2}^{\dagger }|0_{f}\rangle $, $|01\rangle
=f_{3}^{\dagger }|0_{f}\rangle $, $|10\rangle =f_{1}^{\dagger }|0_{f}\rangle 
$, and $|11\rangle =f_{1}^{\dagger }f_{2}^{\dagger }f_{3}^{\dagger
}|0_{f}\rangle$ ~\cite{NayakBasis}. Here $|0_{f}\rangle $ is the vacuum state for fermionic
modes $f_{i}=(\gamma _{i}^{(L)}-i\gamma _{i}^{(R)})/2$, where $\gamma _{i}^{(L)}$
and $\gamma _{i}^{(R)}$ are two Majorana modes on the $i$-th wire. Note that
in this setup with three wires we encode only two qubits~\cite{Georgiev}. This is because
the braiding preserves fermionic parity and, therefore, all states from the 
computational basis must have the same parity (odd in our case).

The Hadamard gates and the oracle unitaries $U_{g_{i}}$ can be implemented by
noting that the braiding of Majorana modes $\gamma _{i}$ and $\gamma _{j}$
is equivalent to the unitary $U_{ij}=e^{\pi \gamma _{i}\gamma _{j}/4}=(%
\mathds{1}-\gamma _{i}\gamma _{j})/\sqrt{2}$. Then, it follows immediately
that $H\otimes \mathds1=U_{12}U_{23}U_{12}$ and $\mathds{1}\otimes
H=U_{56}U_{45}U_{56}$ for the Hadamard gates acting on the first and the
second qubit, respectively, and $U_{g_{1}}=U_{12}^{2}$, $U_{g_{2}}=U_{34}^{2}
$, and $U_{g_{3}}=U_{56}^{2}$ for the oracle unitaries ($U_{g_{0}}=\mathds{1}
$). As a result, the Deutsch-Josza-algorithm can be realized with $14$
braiding operations. In our case, however, the number of operations can be
reduced to nine: The sequence 
\begin{equation*}
\mathcal{U}_{i}=U_{45}U_{56}U_{23}U_{12}U_{g_{i}}U_{56}U_{45}U_{12}U_{23}
\end{equation*}%
acting on $|00\rangle $ gives $\mathcal{U}_{0}|00\rangle =|00\rangle $ for
the constant case and $\mathcal{U}_{1}|00\rangle =i|10\rangle $, $\mathcal{U}%
_{2}|00\rangle =|11\rangle $, and $\mathcal{U}_{3}|00\rangle =i|01\rangle $
for the balanced case. Note also that this protocol can be implemented in
five steps because operations on the Majorana modes $\gamma _{1,2,3}$ and $%
\gamma _{4,5,6}$ before and after the oracle unitary $U_{g_{i}}$ can be
performed in parallel. The final state of the system and, therefore, the
probability to find it in the state $|00\rangle $, can be determined by
measuring the parities of the individual wires in a spectroscopic experiment
\cite{MajoranaDetection} or fermionic number counting \cite{SinglesiteMicroscope}. Taking into account the discussed
insensitivity of the braiding to experimental imperfections, the proposed
protocol provides a robust implementation of the Deutsch-Josza algorithm.

\emph{Conclusion.}  We have presented  an efficient way of braiding MFs
in a cold-atom setup and used it to implement the Deutsch-Josza algorithm in a topologically protected way. 
By adding well-controlled though topologically unprotected operations (e.g. the SWAP-gate), one can go beyond
the braid group and provide a universal "hybrid" set of gates for quantum computation (see also Ref.~\cite{SauHybrid, dasSarmahybrid}). We address this issue in our future work.

\emph{Acknowledgments.} We thank I. Bloch, F. Gerbier, N. Goldman, C. Gross, C. Laflamme, S. Nascimb{\`e}ne and N. Yao for useful comments and discussions. This work has been supported by the Austrian Science Fund FWF (SFB FOQUS F4015-N16),  the US Army Research Office with funding from the DARPA OLE program and  the European Commission via the integrated project AQUTE.


\begin{thebibliography}{99}

\bibitem{TQC_Kitaev}
A~Kitaev.
\newblock {\em Ann. Phys.}, 303:230, 2003.

\bibitem{NayakRMP}
C. Nayak, Steven~H. Simon, A. Stern, M. Freedman, and S.
  Das~Sarma.
\newblock {\em Rev. Mod. Phys.}, 80:1083--1159, 2008.

\bibitem{Pachos}
J.K. Pachos, {\em Introduction to Topological Quantum Computation}, (Cambridge University Press, Cambridge, 2012)

\bibitem{DasSarma_TQC1}
S.  Das~Sarma, M. Freedman, and C. Nayak.
\newblock {\em Phys. Rev. Lett.}, 94:166802, 2005.

\bibitem{DasSarma_TQC2}
C. Zhang, V.W. Scarola, S. Tewari, and S.~Das~Sarma.
\newblock {\em Proc. Natl. Acad. Sci. USA}, 104:18415, 2007.

\bibitem{MajoranaExperiment1}
V.~Mourik, K~Zuo, S.~M. Frolov, S.~R. Plissard, E.~P.~A.~M Bakkers, and L.~P
  Kouwenhoven.
\newblock {\em Science}, 336:1003, 2012.

\bibitem{MajoranaExperiment2}
M.~T Deng, C~.L Yu, G.Y Huang, M.~Larson, P.~Caroff, and H.Q. Xu.
\newblock {\em Nano Lett.}, 12:6416, 2012.

\bibitem{MajoranaExperiment3}
A. Das, Y. Ronen, Y.~Most, Y.~Oreg, M.~Heiblum, and H.~Shtrikman.
\newblock {\em Nat. Phys.}, 8:887, 2012.

\bibitem{Alicea_TQC}
J.~{Alicea}, Y.~{Oreg}, G.~{Refael}, F.~{von Oppen}, and M.~P.~A. {Fisher}.
\newblock {\em Nat. Phys.}, 7:412, 2011.

\bibitem{OregHalperin}
B.~I. Halperin, Y. Oreg, A. Stern, G. Refael, J. Alicea, and
  F. von Oppen.
\newblock {\em Phys. Rev. B}, 85:144501, 2012.

\bibitem{Akhmerov}
M.~Burrello, B.~{van Heck}, and A.~R. {Akhmerov}.
\newblock {\em  arXiv:1210.5452}, 2012.

\bibitem{Beenakker_review}
C.~W.~J. {Beenakker}.
\newblock {\em  arXiv:1112.1950 }, 2011.

\bibitem{Hassler}
F.~{Hassler}, A.~R. {Akhmerov}, C.-Y. {Hou}, and C.~W.~J. {Beenakker}.
\newblock {\em New J. Phys.}, 12:125002, 2010.

\bibitem{Flensberg1}
K. Flensberg.
\newblock {\em Phys. Rev. Lett.}, 106:090503, 2011.

\bibitem{Flensberg2}
M. Leijnse and K. Flensberg.
\newblock {\em Phys. Rev. B}, 86:104511, 2012.

\bibitem{Demler}
D. Pekker, C.-Y. Hou, V.~Manucharyan, and E.~Demler.
\newblock {\em  arXiv:1301.3161}, 2013.

\bibitem{Greiner_singlesite}
J.~{Simon}, W.~S. {Bakr}, R.~{Ma}, M.~E. {Tai}, P.~M. {Preiss}, and
  M.~{Greiner}.
\newblock {\em \nat}, 472:307--312, 2011.

\bibitem{SinglesiteMicroscope}
J.~F. {Sherson}, C.~{Weitenberg}, M.~{Endres}, M.~{Cheneau}, I.~{Bloch}, and
  S.~{Kuhr}.
\newblock {\em \nat}, 467:68--72, 2010.

\bibitem{Probing_Liang}
L.~Jiang, T.~Kitagawa, J.~Alicea, A.~R. Akhmerov, D.~Pekker, G.~Refael, J.~I.
  Cirac, E.~Demler, M.~D. Lukin, and P.~Zoller.
\newblock {\em Phys. Rev. Lett.}, 106:220402, 2011.

\bibitem{MajoranaDetection}
C.~V. Kraus, S.~Diehl, M.~A. Baranov, and P~Zoller.
\newblock {\em New J.  Phys.}, 14:113036, 2012.

\bibitem{Nascimbene}
S.~{Nascimb{\`e}ne}.
\newblock {\em ArXiv e-prints}, 2012.

\bibitem{NayakBasis}
C.~Nayak and F.~Wilczek.
\newblock {\em Nucl. Phys. B}, 479:529, 1996.

\bibitem{Werner}
A. Ahlbrecht, L.~S. Georgiev, and R.~F. Werner.
\newblock {\em Phys. Rev. A}, 79:032311, 2009.

\bibitem{DeutschJosza}
D.~Deutsch and R.~Jozsa.
\newblock {\em Proceedings of the Royal Society of London, Series A}, 439:553,
  1992.

\bibitem{Kitaevchain}
A.~Y. Kitaev.
\newblock {\em Physics-Uspekhi}, 44(10S):131, 2001.

\bibitem{Bloch_singlespin}
C.~{Weitenberg}, M.~{Endres}, J.~F. {Sherson}, M.~{Cheneau}, P.~{Schau{\ss}},
  T.~{Fukuhara}, I.~{Bloch}, and S.~{Kuhr}.
\newblock {\em \nat}, 471:319--324, 2011.

\bibitem{Regev}
O.~Regev and L.~Schiff.
\newblock {\em Proc. of ICALP}, 2008.

\bibitem{Georgiev}
L.~S. Georgiev.
\newblock {\em Phys. Rev. B}, 74:235112, 2006.

\bibitem{SauHybrid}
D.~J. Clarke, J.~D. Sau, and S. Tewari.
\newblock {\em Phys. Rev. B}, 84:035120, 2011.

\bibitem{dasSarmahybrid}
J.~D. Sau, S. Tewari, and S.~Das~Sarma.
\newblock {\em Phys. Rev. A}, 82:052322, 2010.

\end{thebibliography}
\end{document}